\documentclass[showkeys,showpacs]{revtex4}
\usepackage{amsmath}
\usepackage{amssymb}
\usepackage{graphicx,color}
\usepackage{enumerate}

\begin{document}
\title{Phantom ``renormalization'' of mass of a quantized electric charge}

\author{Vladimir Dzhunushaliev}
\email{vdzhunus@krsu.edu.kg}
\affiliation{
Institut f\"ur Physik, Universit\"at Oldenburg, Postfach 2503 
D-26111 Oldenburg, Germany ; \\
Institute for Basic Research, 
Eurasian National University, 
Astana, 010008, Kazakhstan; \\
Institute of Physicotechnical Problems and Material Science of the NAS
of the Kyrgyz Republic, 265 a, Chui Street, Bishkek, 720071,  Kyrgyz Republic }

\author{Ilias Kulnazarov}
\affiliation{Dept. Gen. Theor. Phys., Eurasian National University, Astana, 010008, Kazakhstan}

\author{Ratbay Myrzakulov}
\affiliation{Dept. Gen. Theor. Phys., Eurasian National University, Astana, 010008, Kazakhstan
}
\email{cnlpmyra1954@yahoo.com}

\begin{abstract}
A regular wormhole solution in gravity coupled with a phantom scalar and electromagnetic fields is found. The solution exists  for a special choice of the parameter $f$ of the potential term. The mass $m$ of a wormhole filled 
with a phantom and electrostatic fields is calculated. It is shown that close to some point $f_0$ a small value of the mass $m$ is the remainder of two big masses of the phantom and electrostatic fields. The connection with 
the renormalization procedure in quantum filed theory is considered. The connection between Wheeler's idea ``mass without mass'' and renormalization procedure in quantum field theory is discussed. 
\end{abstract}

\date{\today}

\maketitle

\section{Introduction}

One of the greatest problems in physics is an inner electron structure. In
classical and quantum electrodynamics it is considered that electron is a point-like particle. For example, Feynman diagram technique in quantum electrodynamics is based on the assumption that electron is a point particle. Only in this case the calculations based on this technique are valid. This technique is based on that the interaction between electron and photon takes place in a vertex, i.e. in a point. It means that we assume tacitly that electron is a point-like. 

In classical electrodynamics there is a well known the paradox that the mass $m_E$ of the electric field of the electron  has an infinite value
\begin{equation}
	m_E = \frac{1}{c^2} \int \frac{E^2}{8 \pi} dV = 
	\frac{e^2}{2 c^2} \int \limits_0^\infty \frac{dr}{r^2} = \infty
\label{0-10}
\end{equation}
where $e$ is the electron charge, and $E$ is the electric Coulomb field. The integral diverges on the lower limit as one believes that an electron is a point-like charge. 

In quantum electrodynamics there exists a renormalization procedure. In particular, a renormalized electron mass is the remainder between a ``bare'' mass and a ``counter-term''. Both quantities are infinite but their remainder is finite one and it is equal to the observed electron mass. 

At the beginning of last century Einstein have assumed an idea that the electron is not a point-like but is a wormhole. Further this idea was developed by J. Wheeler \cite{wheel1}. He called such structure as ``mass without mass'' and ``charge without charge''. 

Last decade there were observational facts speaking about accelerated expansion of the presented Universe \cite{Perlmutter} \cite{Riess}. The probable explanation of such fact is that the Universe is filled with dark energy. Dark energy has a negative pressure that leads to the antigravity and accelerated expansion of the Universe. One of model of dark energy is a phantom matter \cite{Caldwell:2003vq} \cite{Baum:2006nz} with negative kinetic term. 

In this paper we would like to show now one can bring together these ideas (``mass without mass'', phantom fields and the renormalization procedure from quantum field theory), and to show that this approach is related to an inner structure of an electron. In order to elaborate this idea we consider a wormhole filled with phantom scalar and electric fields. We show that, in some sense, such construction models a mass renormalization of a point-like charge without spin. 

\section{Mass ``renormalization'' on the background of a phantom wormhole}

In this section we will consider the electric field distribution on the
background of a wormhole created by a phantom scalar field. 

Let us consider Maxwell's electrodynamics on the background of
Minkowski spacetime. The classical electron radius $r_e$, is calculated from the definition of the energy of the electric field 
\begin{equation}
	m_e c^2 = \int \frac{E^2}{8 \pi} dV = \frac{e^2}{2} 
	\int \limits_{r_e}^\infty \frac{dr}{r^2} 
\label{2-50}
\end{equation}
where $E$ is the Coulomb field and $r_e$ is  
\begin{equation}
	r_e = \frac{e^2}{2 m_e c^2} \sim 10^{-13} cm. 
\label{2-60}
\end{equation}
The modern experimental data evidence that the electron has not any structure on the scale $r \sim 10^{-13} cm$. 

Now we would like to consider a wormhole in gravity coupling with a phantom
scalar field. The Lagrangian of the system is 
\begin{equation}
	\mathcal L = - \frac{c^3}{16 \pi G} R - 
	\frac{1}{2c} \partial_\mu \phi \partial^\mu \phi .
\label{2-10}
\end{equation}
In Ref. \cite{ArmendarizPicon:2002km} it is shown that in such system there exists the following (probably the simplest) wormhole solution 
\begin{eqnarray}
	ds^2 &=& c^2 dt^2 - dr^2 - \left( r^2 + r_0^2 \right) \left(
		d \theta^2 + \sin \theta d \varphi^2
	\right), 
\label{2-20}\\
	\frac{d \phi}{dr} &=& \frac{1}{\sqrt{4 \pi}} \frac{q}{r^2 + r_0^2},
\label{2-30}\\
	r_0^2 &=& \frac{G q^2}{c^4} .
\label{2-40}
\end{eqnarray}
Let us consider now electrostatic field on the background of the wormhole
\eqref{2-20} - \eqref{2-40}. The Maxwell equation 
\begin{equation}
	\frac{1}{\sqrt{-g}} \frac{\partial}{\partial x^\nu} \left( 
		\sqrt{-g} F^{\mu \nu}
	\right) = 0 
\label{2-62}
\end{equation}
give us the electric field 
\begin{equation}
	E_{wh} = \frac{e}{r^2 + r_0^2}
\label{2-70}
\end{equation}
where $g$ is the determinant of the metric \eqref{2-20}. The mass of the
wormhole \eqref{2-20} includes the masses of electrostatic \eqref{2-70} and
scalar \eqref{2-30} fields  
\begin{equation}
	m_{wh} = \frac{1}{c^2} \int 
		\left[ \left( T_0^0 \right)_\phi + \frac{E^2_{wh}}{8 \pi}	
	\right]dV = \frac{\pi}{4} \frac{e^2}{c^2 r_0} - 
	\frac{\pi}{4} \frac{q^2}{c^2 r_0} = 
	m_{E} + m_{\phi} 
\label{2-82}
\end{equation}
here $T_0^0$ is the energy-momentum tensor for the phantom field $\phi$ 
\begin{equation}
	T_0^0 = - \frac{{\phi'}^2}{2}. 
\label{2-84}
\end{equation}
and $m_{E} = \frac{\pi}{4} \frac{e^2}{c^2 r_0}$, 
$m_{\phi} = - 	\frac{\pi}{4}\frac{q^2}{c^2 r_0}$ are the masses of the
electric and phantom fields respectively. 
 
In order to make the wormhole mass \eqref{2-82} equal to the electron mass 
$m_e = m_{wh}$, it is necessary to have the following size of wormhole mouth 
\begin{equation}
	r_0 = \frac{\pi}{4} \frac{e^2 - q^2}{m_e c^2} .
\label{2-86}
\end{equation}
We see that by choosing $q$ one can make $r_0$ as small as possible. If $q = e - \delta$ and $\delta \ll e$ then 
\begin{equation}
	r_0 \approx \frac{\pi}{2} \frac{e}{m_e c^2} \delta \sim \frac{\delta}{e} r_e 
	\ll r_e .
\label{2-89}
\end{equation}

\section{Phantom wormhole filled with electric field}

The calculations show that a wormhole solution created by electric
and massless phantom fields has a \textit{negative} energy. In order to
avoid this problem we add a potential energy term for phantom scalar field to the Lagrangian \eqref{2-10}  
\begin{equation}
	\mathcal L = - \frac{c^3}{16 \pi G} R - \frac{1}{c} \left[ 
	\frac{1}{2} \partial_\mu \tilde \phi \partial^\mu \tilde \phi - V(\tilde \phi)
	\right] - 
	\frac{1}{4} F_{\mu \nu} F^{\mu \nu} 
\label{3-10}
\end{equation}
where the potential energy for the phantom scalar field is  
\begin{equation}
	V(\phi) = \frac{1}{2} \left( \frac{\mu}{\tilde f} \right)^2 \left[
		\left( \tilde f^2 \tilde \phi^2 - 1 \right)^2 - 1
	\right]
\label{3-20}
\end{equation}
and $\mu, f$ are some parameters. The field equations are 
\begin{eqnarray}
	R_\mu^\nu - \frac{1}{2} \delta_\mu^\nu R &=& \varkappa T_\mu^\nu 
\label{3-30}\\
	T_\mu^\nu &=& \left( T_\mu^\nu \right)_\phi + 
	\left( T_\mu^\nu \right)_{em},
\label{3-40}\\
	\left( T_\mu^\nu \right)_\phi &=& - \partial_\mu \phi \partial^\nu \phi - 
	\delta_\mu^\nu \left[ 
		\partial_\alpha \phi \partial^\alpha \phi - V(\phi)
	\right], 
\label{3-50}\\
	\left( T_\mu^\nu \right)_{em} &=& \frac{1}{4 \pi} \left(
		- F_{\mu \alpha} F^{\nu \alpha} - \frac{1}{4} \delta_\mu^\nu F_{\alpha \beta} F^{\alpha \beta}
	\right)
\label{3-60}\\
	\frac{1}{\sqrt{-g}} \frac{\partial}{\partial x^\nu} \left( 
		\sqrt{-g} F^{\mu \nu}
	\right) &=& 0 , 
\label{3-64}\\
 	\Box \phi &=& - \frac{d V}{d \phi} 
\label{3-68}
\end{eqnarray}
where $\left( T_\mu^\nu \right)_{\phi, em}$ are the energy-momentum tensors for the phantom scalar and electromagnetic fields, respectively, and $\varkappa = 8 \pi G/c^4$. 

We search a spherically symmetric wormhole solution with the metric
\begin{equation}
	ds^2 = e^{2 \nu(r)} dt^2 - \frac{dr^2}{A(r)} - \left( r^2 + r_0^2 \right) 
	\left(
	d \theta^2 + \sin^2 \theta d \varphi^2
	\right).
\label{3-70}
\end{equation}
After substitution into the Einstein equations \eqref{3-30}, we have the following $\left (^t_t\right )$ and 
$\left (^t_t \right ) - \left (_r^r\right )$ componenets of the Einstein equations
\begin{eqnarray}
	-x \frac{A'}{A} + \frac{1}{A} - 2 + \frac{x^2}{x^2 + x_0^2} &=& 
	\frac{1}{f^2} \frac{x^2 + x_0^2}{A} \left\{
		- A \frac{{\phi'}^2}{2} + \frac{1}{2} \left[
			\left( \phi^2 - 1 \right)^2 - 1
		\right] + \frac{q^2}{8 \pi (x^2 + x_0^2)^2}
	\right\} , 
\label{3-80}\\
	-x \frac{A'}{A} - 2 + \frac{2 x^2}{x^2 + x_0^2} + 2 x \nu' &=& 
	- \frac{1}{f^2} (x^2 + x_0^2) {\phi'}^2 
\label{3-90}
\end{eqnarray}
here $d(\cdots)/dx = (\cdots)'$; $x = \mu r$; $\tilde f \tilde \phi = \phi$; 
$\tilde f^2/\varkappa = f^2$; $\tilde q \tilde f = q$; $x_0 = \mu r_0$ and the
solution of the Maxwell equation \eqref{3-64} is 
\begin{equation}
	F^{01} = e^{-\nu} \sqrt{A}\frac{ \tilde q}{r^2 + r_0^2}.
\label{3-100}
\end{equation}
The field equation for the spherical symmetric phantom field $\phi$ is 
\begin{equation}
	\phi'' + \left( \nu' + \frac{2x}{x^2 + x_0^2} + \frac{1}{2} \frac{A'}{A} \right) = 
	\frac{2}{A} \phi (1 - \phi^2).
\label{3-110}
\end{equation}
The functions $\nu(r), A(r)$ are even functions and $\phi(r)$ is odd function 
\begin{equation}
	\nu'(0) = A'(0) = \phi(0) = 0.
\label{3-120}
\end{equation}
Taking into account \eqref{3-120} we find the following constraints
\begin{equation}
	\phi'(0) = \sqrt{2}\, \frac{f}{x_0}, \quad 
	A(0) = 1 - \frac{x_0^2}{f^2} \left\{
		\frac{1}{2} \left[  
			\left( \phi^2 - 1 \right)^2 - 1 
		\right] + \frac{q^2}{8 \pi x_0^4}
	\right\}. 
\label{3-140}
\end{equation}
Thus the boundary conditions we choose in the form 
\begin{eqnarray}
	\nu(0) &=& 0 , 
\label{3-150}\\
	\phi(0) &=& 0, \quad \phi'(0) = \sqrt{2}\, \frac{f}{x_0},
\label{3-160}\\
	A(0) &=& 1 - \frac{x_0^2}{f^2} \left\{
		\frac{1}{2} \left[  
			\left( \phi^2 - 1 \right)^2 - 1 
		\right] + \frac{q^2}{8 \pi x_0^4}
	\right\}. 
\label{3-170}
\end{eqnarray}
The condition \eqref{3-150} follows from the fact that we can choose arbitrary the time scale in the metric \eqref{3-70}. We will seek for asymptotically flat solutions. 

\section{Mass ``renormalization''}

\begin{figure}[h]
\begin{minipage}[t]{.45\linewidth}
 \begin{center}
 \fbox{
  	\includegraphics[height=.8\linewidth,width=.8\linewidth]{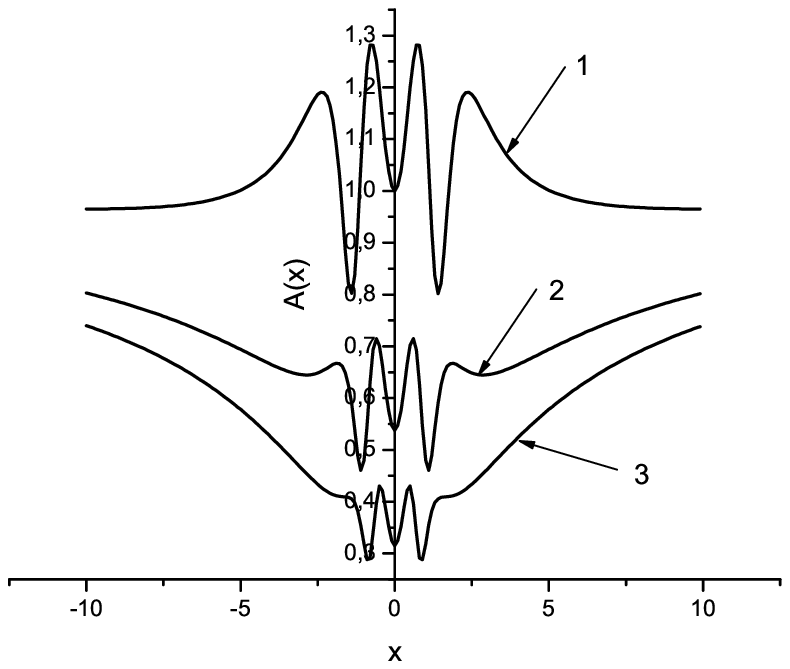}
  }
  \caption{The profiles of the functions $A(x)$. The curves 1,2,3 present $A(x)$ for $q=0,10,20$. 
  The corresponding values of the parameter $f^*$ are $f^* = 2.335454422, 2.486821, 2.9332549$
  }
  \label{fg1}   
	\end{center}
\end{minipage}\hfill
\begin{minipage}[t]{.45\linewidth}
 \begin{center}
 \fbox{
	  \includegraphics[height=.8\linewidth,width=.8\linewidth]{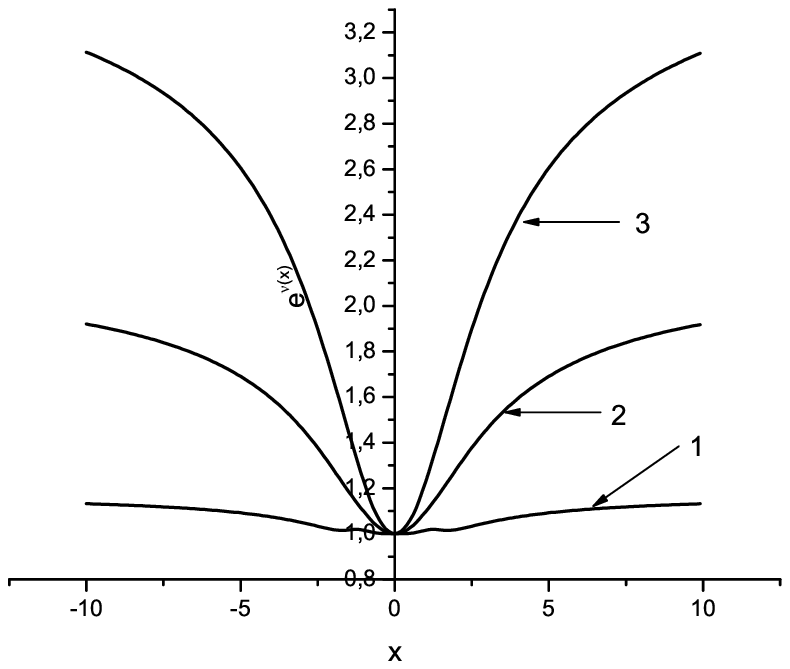}  
 }
	\caption{The profiles of the functions $e^{\nu(x)}$. The curves 1,2,3 present 
	$e^{\nu(x)}$ for $q=0,10,20$. 
  The corresponding values of the parameter $f^*$ are $f^* = 2.335454422, 2.486821, 2.9332549$
	}
	\label{fg2}   
	\end{center}
\end{minipage}\hfill 
\end{figure}

\begin{figure}[h]
\begin{minipage}[t]{.45\linewidth}
 \begin{center}
 \fbox{
  	\includegraphics[height=.8\linewidth,width=.8\linewidth]{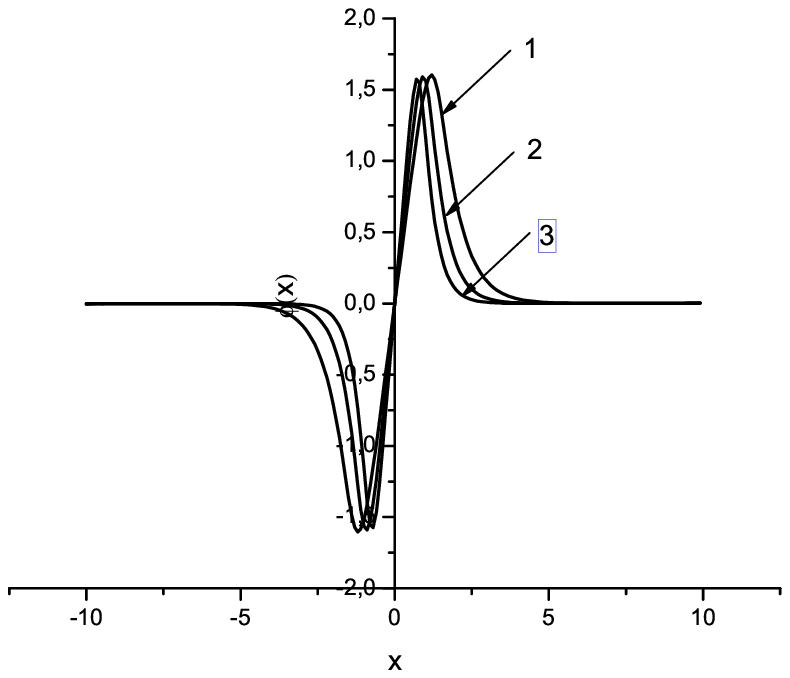}
  }
  \caption{The profiles of the functions $\phi(x)$. The curves 1,2,3 present $\phi(x)$ for $q=0,10,20$. 
  The corresponding values of the parameter $f^*$ are $f^* = 2.335454422, 2.486821, 2.9332549$
  }
  \label{fg3}   
	\end{center}
\end{minipage}\hfill
\begin{minipage}[t]{.45\linewidth}
 \begin{center}
 \fbox{
	  \includegraphics[height=.8\linewidth,width=.8\linewidth]{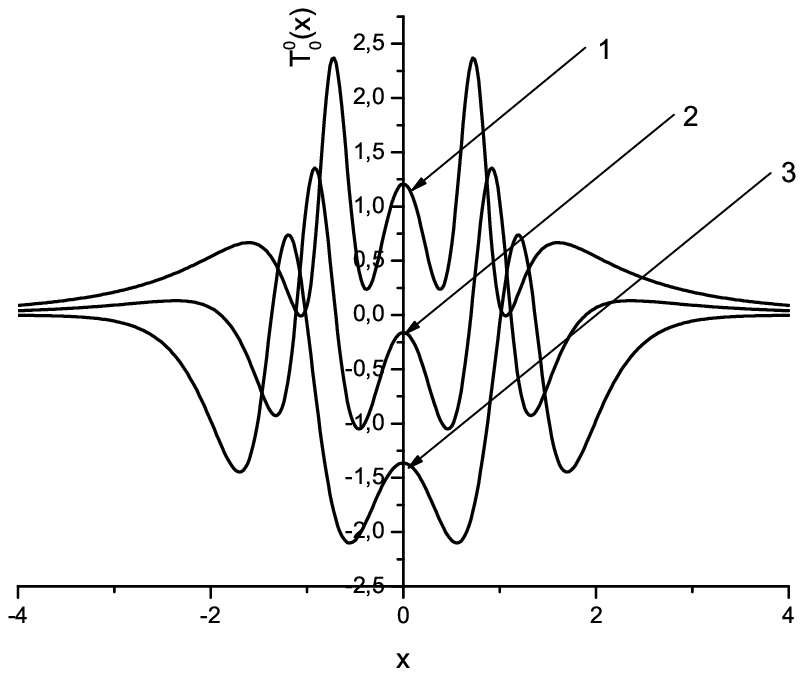}  
 }
	\caption{The profiles of the dimensionless energy density $T^0_0(x)$. The curves 1,2,3 present 
	$T^0_0(x)$ for $q=0,10,20$. 
  The corresponding values of the parameter $f^*$ are $f^* = 2.335454422, 2.486821, 2.9332549$
	}
	\label{fg4}   
	\end{center}
\end{minipage}\hfill 
\end{figure}

The mass of an asymptotically flat solution is given by the formula  
\begin{equation}
\begin{split}
	\tilde m = &\frac{4 \pi}{c^2} \int\limits_0^\infty T^0_0 (r^2 + r_0^2) dr = 
	\frac{4 \pi}{\mu \tilde f^2 c^2} \int\limits_0^\infty \left\{
		- A {\phi'}^2 + \frac{1}{2} \left[
			\left( \phi^2 - 1 \right)^2 - 1 
		\right] + \frac{\tilde q^2}{8 \pi \left( x^2 + x_0^2 \right)^2}
	\right\} \left( x^2 + x_0^2 \right) dx = 
\\
	&
	m_{\phi} + m_E = 
	\frac{4 \pi}{\mu \tilde f^2 c^2} m(f^*) = 
	\frac{c^2}{2 G} \frac{1}{\mu} \frac{m(f^*)}{{f^*}^2}
\label{3-180}
\end{split}
\end{equation}
where 
\begin{equation}
	m_{\phi} = \frac{4 \pi}{c^2} \int\limits_0^\infty 
	\left( T^0_0 \right)_\phi (r^2 + r_0^2) dr < 0  
\label{3-190}
\end{equation}
is the mass of the phantom field and 
\begin{equation}
	m_{E} = \frac{4 \pi}{c^2} \int\limits_0^\infty 
	\left( T^0_0 \right)_E (r^2 + r_0^2) dr > 0  
\label{3-200}
\end{equation}
is the mass of the electrostatic field 
\begin{equation}
	E_r = \sqrt{\left| 
		F_{01} F^{01}
	\right| } = 
	\dfrac{{\tilde q}^2}{r^2 + r^2_0}.
\label{3-210}
\end{equation}
The numerical analysis shows that regular solutions exist only for a special choice of the parameter $f=f^*$. This means that we have a nonlinear eigenvalue problem. The results of numerical calculations \ref{fg1}-\ref{fg4} are presented in Fig's. 

\begin{figure}[h]
 \begin{center}
 \fbox{
  	\includegraphics[height=.5\linewidth,width=.5\linewidth]{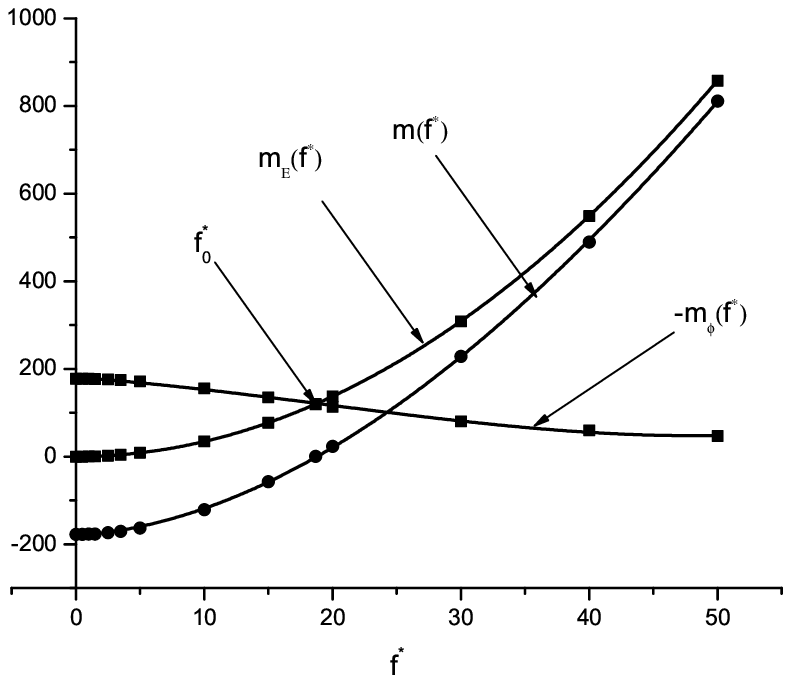}
  }
  \caption{The profiles of $m_E(f^*), (-m_\phi(f^*))$ and $ m(f^*)$.
  }
  \label{fg5}   
	\end{center}
\end{figure}

In Fig. \ref{fg5} the profile of $m(f^*)$ is presented. In Table \ref{tbl1} the values of $q, f^*$ and $m(f^*)$ are presented. 

\begin{table}
\begin{center}
\caption{
}
\label{tbl1}
\begin{tabular}{|c|c|c|c|c|} \hline
	electric charge $q$	&	eigenvalue $f^*$	&dimensionless mass $m_\phi(f^*)$	&dimensionless mass $m_E(f^*)$	&	dimensionless mass $m(f^*)$	\\	\hline
	.0						&	2.335454422	&-177.6721052266339	&	0
			&	-177.67210527336235					\\	\hline
	.5		&	 2.335832		&-177.6106087127155	&	0.08580629762419899	
			&	-177.52480247076312					\\	\hline
	1.		&	2.336965		&	-177.4259691322357	& 0.34322519049679595 &
-177.08274391936118	\\	\hline
	1.5		&	2.338855		&	-177.18318129138063	& 0.7722566786177909 &
-176.41092455722256					\\	\hline
	2.5		&	2.344898	&	-176.14594505461523`	&	2.145157440604975	&
-174.0007757913777					\\	\hline
	3.5		&	2.353967	&	-174.70628724455838	&	4.204508583585751	&-170.50177868299394					\\	\hline
	5.		&	2.373252	&-171.70634752583763	&	8.580629762419898
&	-163.12571787930372					\\	\hline
	10.			&	2.486821	&-155.82229033094694&34.3225190496796	&	-121.49977116200269					\\	\hline
	15.		&	2.675175	&-134.8786363719572	&77.22566786177909&	 -57.65296826750314					\\	\hline
	18.690326360554604	&	2.8594037	&-119.11167254238663&119.8982721294677				&	0.7865997856544097					\\	\hline
	20.		&	2.9332549	&-113.72372268940026&137.29007619871837				&	23.56635426688145						\\	\hline
	30.		&	3.61553334	&-80.27042545281925&308.9026714471164			&	228.6322450558395						\\	\hline
	40.		&	4.438410324	&-59.43611889669617&549.1603047948735			&	489.7241841555605						\\	\hline
	50.		&	5.333364895	&-46.58773074475898&858.0629762419899			&	811.4752335617483						\\	\hline
\end{tabular}
\end{center}
\end{table}

One can find a fitting curve for the points $m(f^*)$ and $m_{\phi, E}$
\begin{eqnarray}
	m(f^*) &=&  -180.491 + 2.03601 x + 0.431101 x^2 - 0.00151353 x^3, 
\label{3-220}\\
	m_\phi(f^*) &=&  -180.491 + 2.03601 x + 0.0878758 x^2 - 0.00151353 x^3, 
\label{3-222}\\
	m_E(f^*) &=&  0.343225 x^2 .
\label{3-224}
\end{eqnarray}
One can see from Fig. \ref{fg5} that there exists a point $f_0^*$ where
$m(f^0_*) = m_E(f^*_0) - | m_\phi(f^*_0)| = 0$. This means that close to this
point one can model the renormalization procedure in the sense that the ratio
$(m_{E} - |m_\phi|)/ m_E$ can be made arbitrary small 
\begin{equation}
	\frac{m_{E} - |m_\phi|}{m_E} \ll 1
\label{3-230}
\end{equation}
when 
\begin{equation}
	\left |
		\frac{f^* - f_0^*}{f_0^*}
	\right | \ll 1 .
\label{3-240}
\end{equation}
If we would like to have $\tilde m = m_e$ and choose $x_0= \mu r = 1$ then 
\begin{equation}
	\frac{m (f^*)}{{f^*}^2} = \frac{2 G}{c^2} \mu m_e = 10^{-54} \mu .
\label{3-250}
\end{equation}
If we take a minimal length in Nature, $1/\mu \sim 10^{-33}$cm, then 
\begin{equation}
	\frac{m(f^*)}{{f^*}^2} \sim 10^{-21} .
\label{3-260}
\end{equation}
We see from Fig. \ref{fg5} that at this point $- m_\phi \sim m_E \sim 10^2$, i.e. that 
\begin{equation}
	\frac{|m_\phi (f^*)|}{m(f^*)} \sim \frac{|m_E(f^*)|}{m(f^*)} \sim 
	10^{21} \gg 1 .
\label{3-270}
\end{equation}
It happens in quantum field theory that a bare mass $m_{b}$ and a renormalization mass $m_r$ are infinite 
($m_{b}= \infty$ and $m_{r}= \infty$) but the remainder between them is finite 
\begin{equation}
	m_{b} - m_{r} = m_e .
\label{3-280}
\end{equation}
where $m_e$ is the observable electron mass. In our situation we have $m_{\phi, E} \neq \infty$ but 
\begin{equation}
	m_{\phi, E} \gg m_E - | m_\phi | .
\label{3-290}
\end{equation}
Another interesting feature of a phantom regularization considered here is that we have one to one correspondence between the mass $m(f^*)$ and the charge $\tilde q$: for every $q$ there exists an exceptional value of $f^*$ and correspondingly only one value of the mass $m(f^*)$. One can say that the electric charge is quantized by the phantom regularization of the mass of the  wormhole filled with the electric charge.

\section{Discussion and conclusions}

We have shown that in gravity + phantom scalar field + electromagnetism there exists a regular wormhole solution for a special choice of parameters, i.e. the problem of finding of regular solutions in this approach is a nonlinear eigenvalue problem. In this approach there exists one to one correspondence between the electric charge and the mass of a wormhole. For every pair there exists only one regular wormhole solution for a special choice of the parameter $f$ of the potential energy of the phantom scalar field. 

On the basis of calculations presented here we can hypothesize that if an  electron has an inner structure (which can be explained in quantum gravity) then a phantom scalar field (in an effective manner) models (mimics)  the renormalization procedure in quantum field theory. 

Thus here we bring together three ideas: ``mass without mass'', phantom fields
and renormalization procedure from quantum field theory. The result is: probably
in quantum gravity these ideas can be connected and give us an inner structure
of the electron.  

\section*{Acknowledgements}

V.D. is grateful to the Research Group Linkage Programme of the Alexander von Humboldt Foundation for the support of this research and would like to express the gratitude to the Department of Physics of the Carl von Ossietzky University of Oldenburg  and, specially, to V. Folomeev, J. Kunz and B. Kleihaus for fruitful discussions.

\end{document}